\documentclass[showpacs,twocolumn,prb]{revtex4}
\usepackage{graphicx}
\usepackage{epsfig}
\usepackage{amssymb}

\def\beq{\begin{equation}}
\def\eeq{\end{equation}}
\def\beqa{\begin{eqnarray}}
\def\eeqa{\end{eqnarray}}

\def\e{\epsilon}

\def\half{{\ss 1\over 2}}

\def\D{\Delta}
\def\del{\delta}
\def\e{\epsilon}
\def\cH{{\mathcal H}}
\def\cL{{\mathcal L}}
\def\ss{\scriptstyle}

\def\Tr{\mathrm{Tr}}
\def\r{{\bf r}}

\def\ss{\scriptstyle}

\def\si{\sigma}
\def\etal{{\sl et.al.}}

\def\cdag{c^{\dagger}}

\def\nonum{\nonumber \\}

\def\del{\delta}

\def\nonum{ \nonumber \\}

\newcommand{\bra}[1]{| #1 \rangle}
\newcommand{\ket}[1]{\langle #1 |}

\begin{document}

\title{Effective single-particle order-$N$ scheme for the dynamics of
open non-interacting many-body systems}
\author{Yu. V. Pershin, Y. Dubi and M. Di Ventra}
\affiliation{Department of Physics, University of California San
Diego, La Jolla, California 92093-0319, USA}

\begin{abstract}

Quantum master equations are common tools to describe the dynamics
of many-body systems open to an environment. Due to the interaction
with the latter, even for the case of non-interacting electrons, the
computational cost to solve these equations increases exponentially
with the particle number. We propose a simple scheme, that allows to
study the dynamics of $N$ non-interacting electrons taking into
account both dissipation effects and Fermi statistics, with a
computational cost that scales linearly with $N$. Our method is
based on a mapping of the many-body system to a specific set of
effective single-particle systems. We provide detailed numerical
results showing excellent agreement between the effective
single-particle scheme and the exact many-body one, as obtained from
studying the dynamics of two different systems. In the first, we
study optically-induced currents in quantum rings at zero
temperature, and in the second we study a linear chain coupled at
its ends to two thermal baths with different (finite) temperatures.
In addition, we give an analytical justification for our method,
based on an exact averaging over the many-body states of the
original master equations.
\end{abstract}

 \maketitle

\section{Introduction}
  Quantum systems that exchange energy with an environment have
attracted a great deal of attention for many years
\cite{Feynman,Legget}. The interest in these dissipative (open)
quantum systems ranges from quantum computing and quantum
information theory to biological physics~\cite{Weiss}. Recent
developments in the transport properties of nanoscale systems
\cite{Max} raise new interest in these topics. For instance, the
dissipative effects of the surrounding environment are key to
understand the non-equilibrium properties of nanostructures and
their approach to steady state~\cite{Bush1}. However, the study of
dissipative many-body quantum systems represents a major
computational challenge.

There are essentially two ways to approach this problem. One
consists in deriving equations of motion (master equations) for the
reduced density matrix (DM) of the system of interest by integrating
out the degrees of freedom of the bath.~\cite{Redfield} The further
assumption of Markovian dynamics leads to different kinds of master
equations for the DM~\cite{vancamp}, perhaps the most popular being
the Lindblad equation~\cite{lind} which is often used in quantum
optics~\cite{Louisell,Breuer}. The second approach is to use stochastic
Schr\"odinger equations~\cite{vancamp,Breuer} which are the stochastic
unraveling of the master equations. If the Hamiltonian of the system
does not depend on microscopic degrees of freedom, like the density
or current density, both approaches describe the same physical
properties.~\cite{DiVentra2007}

Irrespective of the chosen method, the solution of these equations
is a formidable task which scales exponentially with the number of
electrons. This is true even for a system of non-interacting
electrons since the correlations induced by the bath make it
impossible to exactly reduce the $N$-particle equation of motion
into $N$ distinct single-particle equations of motion. It is the
goal of this paper to discuss an {\em ansatz} which greatly
simplifies this task for the dynamics of $N$ non-interacting
electrons in interaction with a bath. We focus on the DM approach
but the conclusions are exactly the same for the stochastic
Schr\"odinger equations. The latter, in fact, have found application
in the recently developed stochastic time-dependent current-density
functional theory (S-TDCDFT)~\cite{DiVentra2007}, an extension of
time-dependent current-density functional theory to systems in
dynamical interaction with an environment. In S-TDCDFT the many-body
interacting problem in the presence of the environment is mapped
into an effective single-particle non-interacting problem in the
presence of the same environment. The {\em ansatz} we discuss in
this work is thus of great use in the numerical solution of the
equations of motion of S-TDCDFT,~\cite{DiVentra2007} and may
therefore find application in disparate problems beyond the examples
presented in this paper, where interactions are important.

The outline of the paper is as follows. In Sec.~\ref{scheme} we
describe in detail our proposed scheme. We define the master
equation framework and our {\em ansatz}, along with the detailed
structure of the resulting equations. In Secs.~\ref{numeric} and
\ref{numeric1} we give numerical examples of our scheme. We
calculate currents induced by optical excitation in quantum
 ring structures in the presence of dissipation at $T=0$ (Sec.~\ref{numeric}) and consider steady state properties of a
quantum system at finite temperatures (Sec.~\ref{numeric1}). We
study systems which are small enough so that we can compare the
results from our scheme with the full many-body calculation. We find
excellent agreement between the two methods for a large range of
parameters. In Sec.~\ref{analytic} we derive an analytical
justification for our scheme. Starting from the exact many-body
master equations we average over the many-body degrees of freedom
and study the resulting (non-linear) equations. Sec.~\ref{summary}
is devoted to a summary and outlook.

\section{Calculation scheme} \label{scheme} Our goal is to study the dynamics of $N$
electrons described by a non-interacting Hamiltonian
$\cH=\sum_jH_j$, while taking into account dissipation processes. To
be more specific let us employ the following Lindblad-type master
equation for the many-electron DM $\rho_M$ ($\hbar=1,~e=1$)
\cite{lind} \beq \dot{\rho}_M = -i[\cH,\rho_M]+\cL \rho_M ~~,
\label{lind_eq1} \eeq where $[\cdot]$ denotes the commutator, and
$\cL$ is the
 Lindbladian superoperator, defined via a set $V_{nn'}$ of so-called ``Lindblad
 operators'' by
 \beq
 \cL \rho_M = \sum_{n,n'} \left( -\half \{ V^\dagger_{nn'} V_{nn'}, \rho_M \}
 +V_{nn'} \rho_M V^\dagger_{nn'} \right) ~~, \label{Lindbladian}
 \eeq
with $\{\cdot\}$ the anti-commutator. The sums over $n$ and $n'$
($n\neq n'$) are performed over all many-particle levels of the
system and the $V$-operators are conveniently selected in the form
$V_{n n'}=\sqrt{ \gamma_{nn'}} \bra{\Psi_n}\ket{\Psi_{n'}}$,
describing a transition from the many-body state $\bra{\Psi_{n'}}$
into the state $\bra{\Psi_n}$ with the transition rate
$\gamma_{nn'}$. Although $\gamma_{nn'}$ are introduced
phenomenologically here, these coefficients can be in principle
derived from a microscopic theory.

A common form for $\gamma_{nn'}$ is described as follows
\cite{Neil}. At $T=0$, dissipation drives the system towards its
ground state, which we denote by the index $n=1$. Therefore, it is
reasonable to select $\gamma_{nn'}=0$ for $n>1$. Moreover, by
assuming that the transition rate into the ground state is
independent of $n'$ , we may write $\gamma_{1,n'}=\gamma$. This
choice for the relaxation rates is a $T=0$ manifestation of detailed
balance \cite{vancamp}, which we assume to hold for a Markovian
ohmic  bath in the long-time limit. In fact, there are other ways to
choose the relaxation operators and still ensure detailed balance,
and we have checked different options in our numerical calculations
(Sec.~\ref{numeric}) and found no qualitative change in our results.
Therefore, we shall keep the above normalization hereon.

 For a system with $M$ single-electron
energy levels and $N$ electrons, the solution of
Eq.~(\ref{lind_eq1}) generally requires the solution of
$(C^M_N+2)\times(C^M_N-1)/2$ coupled differential equations, where
$C^M_N=M!/N!(M-N)!$ and we have taken into account constrains of
hermiticity and the unit trace of the density matrix. For the
general case (excluding, e.g., $N=1$ or $N=M$), the problem thus
scales exponentially with the number of particles \cite{stochN}.

Consider now an operator $A=\sum_j A_j$, a sum over single-particle
operators. (This is not the most general form of operator but it
encompasses most of the observables of physical interest, like,
e.g., the density or current density.) We make the following
conjecture: the expectation value of $A$ over a many-particle
non-interacting electron state with dissipation can be approximated
as a sum of single-electron expectation values of $A_j$ over an
ensemble of $N$ single-electron systems with {\em specifically
selected} single-electron dissipation operators, i.e.
\begin{equation}
\textnormal{Tr}A\rho_M\simeq
\sum\limits_{j=1}^N\textnormal{Tr}A_j\rho^{(j)}. \label{theorem}
\end{equation}
Here, $\rho^{(j)}$ is a single-electron DM (effectively describing
the $j$-th electron), each obeying its own Lindblad master
equation \beq \dot{\rho}^{(j)} = -i[H_j,\rho^{(j)}]+\cL^{(j)}
\rho^{(j)} ~~. \label{lind_eq2} \eeq The choice of superoperators
$\cL^{(j)}$ is dictated by two requirements: (i) for a
time-independent Hamiltonian the dissipation processes should
result in the Fermi-Dirac distribution at long times, and (ii) the
relaxation rate of many-electron states is $\gamma$.

As we will demonstrate (numerically in Sec.~\ref{numeric} as well as
analytically in Sec.~\ref{analytic}), these two requirements are met
if one chooses a simple form for the $V$-operators, which reflects
the physical process at which the different electrons decay to
consecutive single-particle levels (i.e., the $i-$th electron will
decay to the $i$-th single-particle level, see
Eq.~(\ref{V_effective})). Once a form for $\cL^{(j)}$ is chosen, one
only needs to solve $\sim N\times M^2$ equations, a reduction which
enormously speeds up numerical calculations.

The simplest choice for the Lindbladian superoperator which
satisfies the above criteria is similar to the one in Eq.
(\ref{Lindbladian}), with single-electron $V$ operators of the
following form: for the $j$-th electron we select at $T=0$ \beq
V^j_{kk'}=\left\{
\begin{array}{cl}
  \sqrt{\gamma} \bra{j}\ket{k'}&,~~k' \neq  k= j;~~ k \leq k_F \\
  0&,~~\mathrm{otherwise}
\end{array} \right.,
\label{V_effective} \eeq where $\bra{k}$ are now the single-particle
states and $k_F$ is the index of the Fermi level. In some sense,
such a replacement of the many-body equation of motion by a set of
auxiliary single-electron equations is similar to the introduction
of a fictitious system of non-interacting electrons in
density-functional theory.~\cite{DiVentra2007}

To summarize our scheme, it is constructed from the following steps:
(i) given a non-interacting Hamiltonian, one constructs a set of
Lindblad operators (following Eq.~(\ref{Lindbladian}) and
Eq.~(\ref{V_effective})), (ii) a set of single-particle density
matrices ${\rho}^{(j)}$ is defined, and corresponding master
equations [Eq.~(\ref{lind_eq2})] are solved, and, finally, (iii) any
observable quantity (made of quadratic operators in the second
quantization formalism) can be calculated using Eq.~(\ref{theorem}).

\section{Numerical demonstration: driven system at $T=0$}\label{numeric}  In order to test the conjecture
(\ref{theorem}), we have performed extensive numerical
calculations considering a driven quantum system in a wide range
of parameters. We found that for a system with non-degenerate
levels Eq. (\ref{theorem}) is almost perfectly satisfied. We
believe that in systems with degenerate energy levels a deviation
from Eq. (\ref{theorem}) is due to the intrinsic ambiguity of
degenerate states.

\begin{figure}[t]
\includegraphics[width=8truecm]{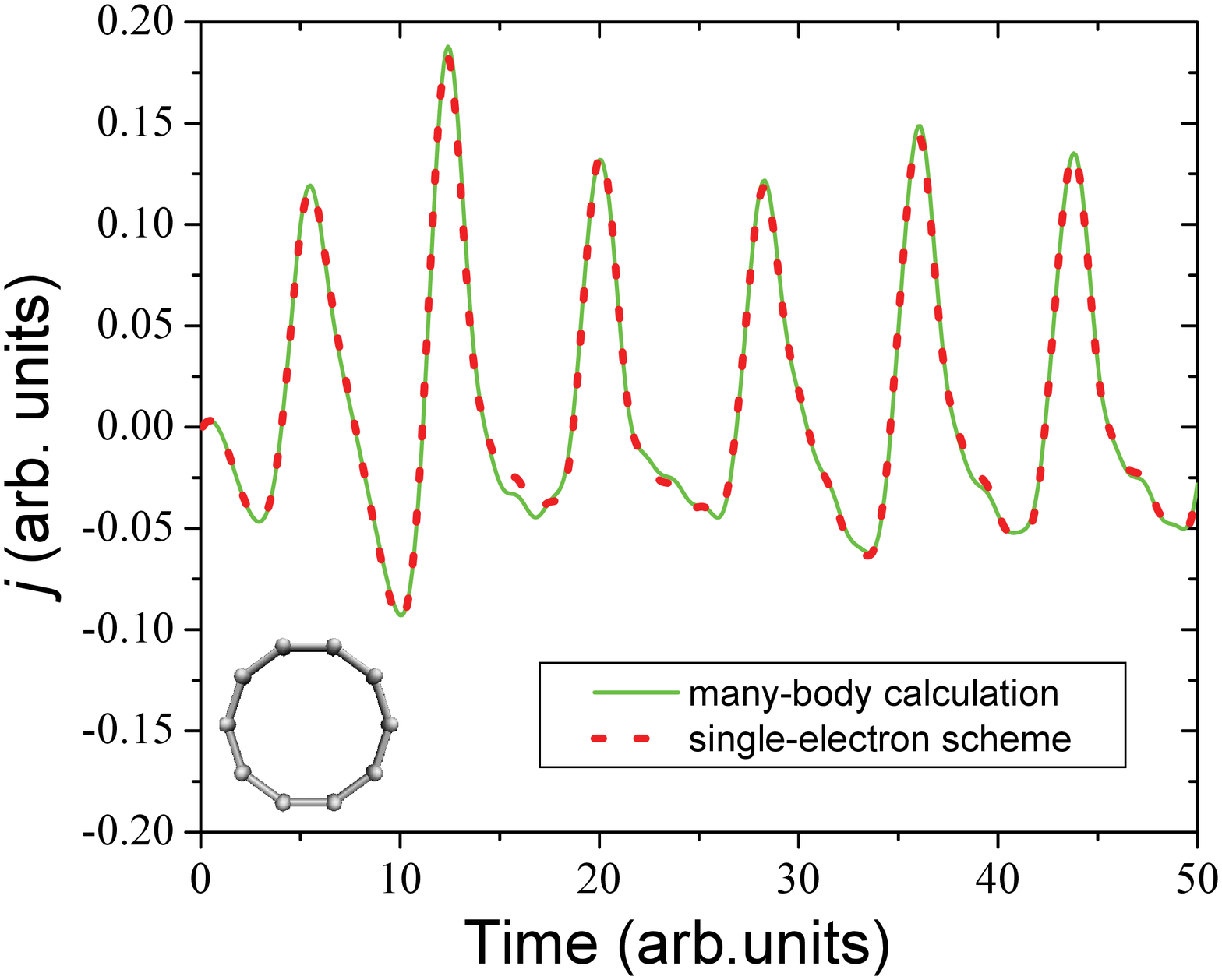}
\caption{\label{fig1} (Color online) Current between two sites of
a ring as a function of time calculated by the exact many-body and
approximate single-electron approaches. Inset shows the system
geometry. This calculation has been done with the following set of
parameters: $N=3$, $M=10$, $eE_0a=0.1$, $\omega=0.8$, $\si= 1$,
$\gamma=0.1$, $a=0.1415$nm and $B=10$T. $a$ is a bond length. The
magnetic field corresponds to a flux through the ring of
$\phi/\phi_0 \approx 1.66\times 10^{-4}$.}
\end{figure}

We consider a
system of $N$ tight-binding electrons on both a ring and a double
ring of $M$ sites in the presence of circularly-polarized
electromagnetic radiation (see insets in Fig.~\ref{fig1}). In order
to lift the degeneracy, we place the system in a weak magnetic flux.
The Hamiltonian of the system is given by \beq \cH=- t\sum_{i}
\left( e^{i 2\pi \phi/\phi_0}\cdag_i c_{i+1}+h.c. \right) +\sum_{i}
U_{i}(t) \cdag_i c_i~~. \label{Ham} \eeq Here $t$ is the hopping
integral (we set $|t|=1$ throughout the calculation) and
$U_i(t)=-e\mathbf{E}(t)\cdot \mathbf{r}_i$ is a change of the
potential energy of the $i$-th site ($\r_i$ is its position) due to
the external radiation. The magnetic field is taken into account via
the usual Peierls substitution, with $\phi$ the magnetic flux
through the ring, and $\phi_0=h/e$ the flux quantum. The electric
field is written as $\mathbf{E}(t)=E_0\cos (\omega t){\bf
\hat{x}}\pm E_0\sin (\omega t) {\bf \hat{y}}$, where $E_0$ and
$\omega$ are the electric field amplitude and frequency, ${\bf
\hat{x}}$ and ${\bf \hat{y}}$ are unit vectors in the $x$ and $y$
directions (in the ring plane), and $\pm$ corresponds to a
$\sigma_\pm$ circular polarization.

It is known that in the ring topology a circularly-polarized
radiation creates a current in the ring \cite{perpier,Nobusada}. We
calculate the expectation value of the current operator through a
specific bond, $J=\frac{ie}{\hbar}\langle \cdag_i
c_{i+1}-h.c.\rangle$, using both the exact many-body DM, and a set
of single-electron density matrices calculated as described
above.~\cite{prec} In both schemes we start by diagonalizing the
tight-binding part of the Hamiltonian. In the many-body (exact)
scheme, we then write the time-dependent potential and the Lindblad
operators in their full many-body form and solve the time-dependent
set of equations for the many-body
DM. For the single-particle scheme, we solve a set of $N$
single-particle Lindblad equations (of size $M \times M$), each with
its own set of relaxation operators $\cL^{(j)}$. The current is then
calculated as a function of time using the LHS (many-body form) and
the RHS (single-particle form) of Eq.~(\ref{theorem}). The
calculations were made for a wide range of system parameters,
displaying excellent agreement between the two schemes.

\begin{figure}[tb]
\includegraphics[width=8truecm]{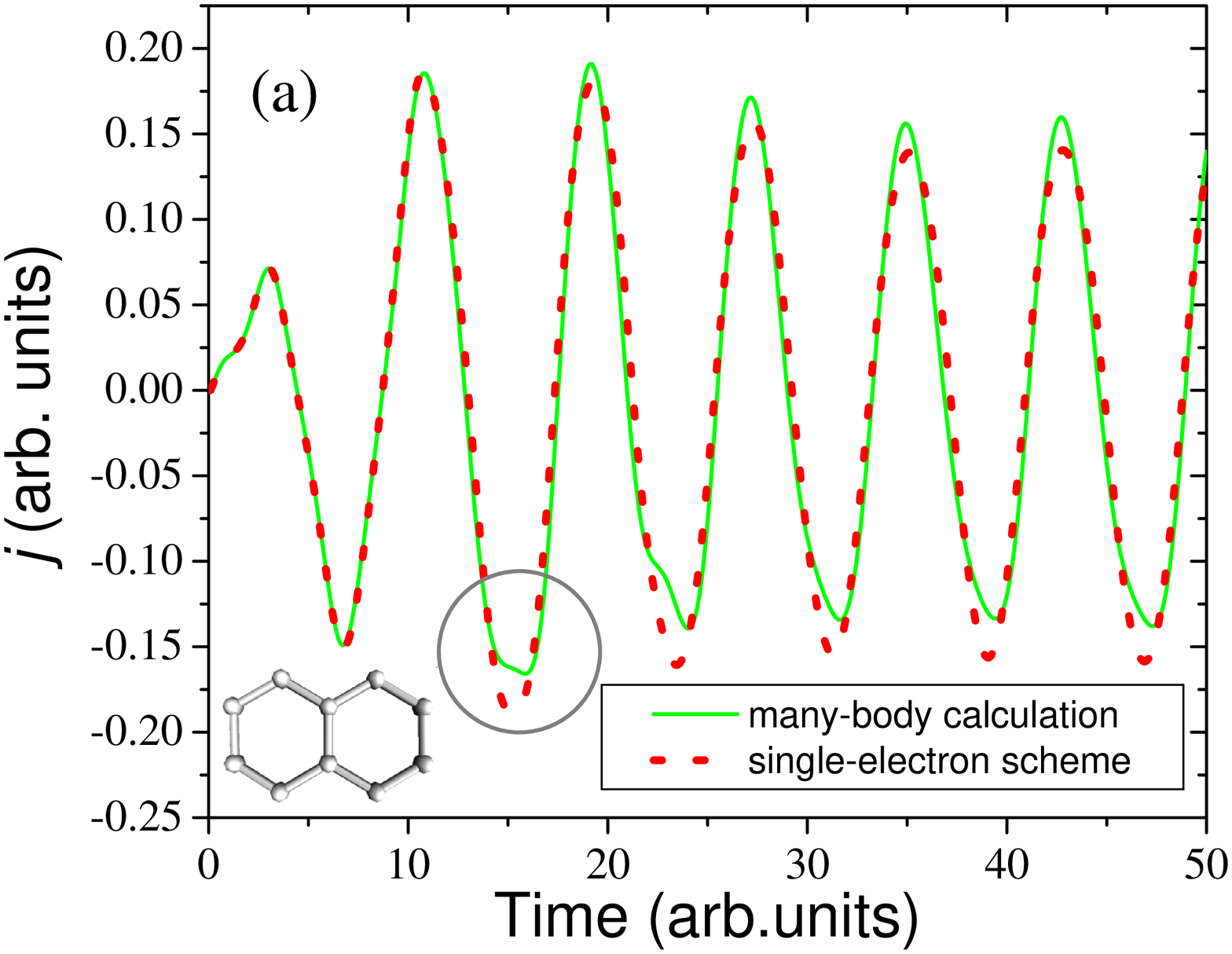}
\includegraphics[width=8truecm]{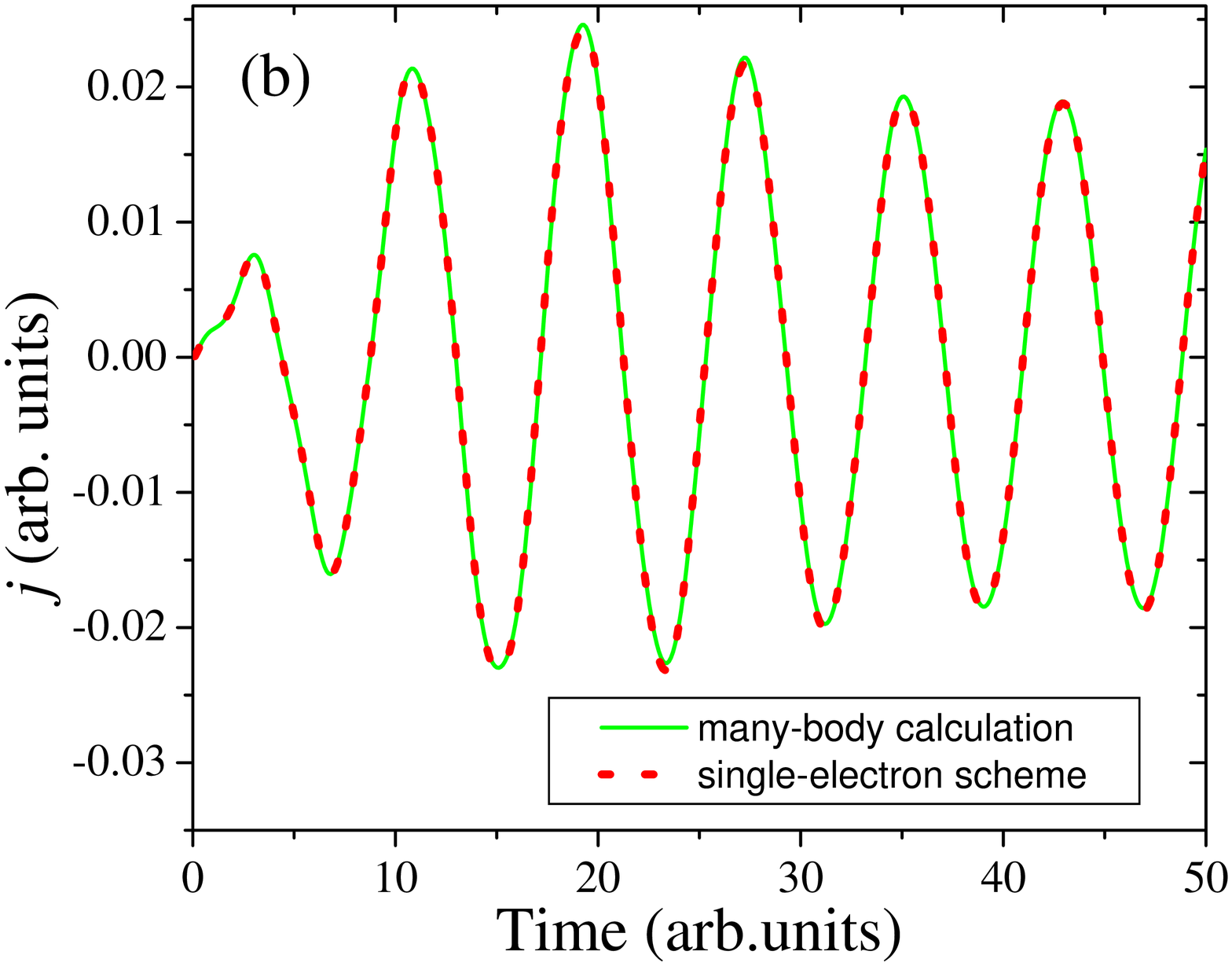}
\caption{\label{fig2} (Color online) Current excited in a double
ring calculated by the exact many-body and approximate
single-electron approaches. The electric field amplitude is (a)
$eE_0a=0.1$ and (b) $eE_0a=0.01$. All other parameters are the
same as in Fig. \ref{fig1}. The system geometry is shown in the
inset of (a). The discrepancies between the two methods (such as
marked in a gray circle in (a)) diminish as the excitation field
decreases.}
\end{figure}

\subsection{Ground-state initial conditions}

Fig.~\ref{fig1} shows the current calculated by the two methods
through a bond connecting two adjacent sites of a 10-site ring
containing 3 electrons.  We see that the current through the bond
oscillates in agreement with a previous study \cite{Nobusada}. Most
importantly, in the context of the present investigation, the
current values hardly differ between the two schemes. The average
deviation of the two currents is less then $1.5 \%$ (the maximum
deviation is $\approx6.5 \%$). This difference rapidly disappears
with decreasing $E_0$. This is seen from comparing
Fig.~\ref{fig2}(a) and \ref{fig2}(b) where the current excited in a
double ring structure is plotted for two different values of the
electric field amplitude, $eE_0a=0.1$ and $eE_0a=0.01$,
respectively. One clearly sees that the discrepancies between the
two methods (marked in a gray circle in \ref{fig2}(a)) diminish as
the excitation field decreases. The results presented in Figs.
\ref{fig1} and \ref{fig2} were obtained assuming that at time $t=0$
the system is in its ground state.

\subsection{Non-equilibrium initial conditions}

\begin{figure}[bt]
\includegraphics[width=8truecm]{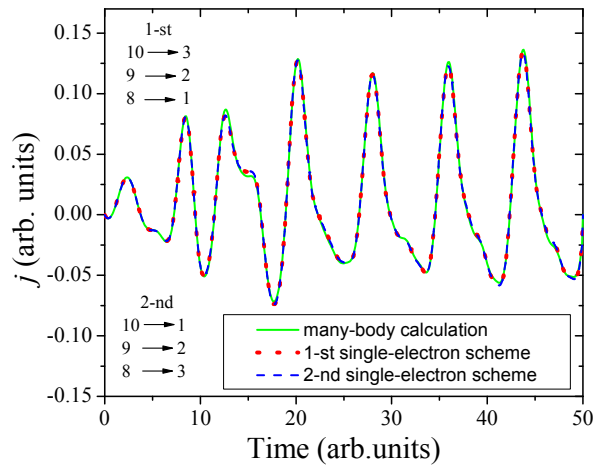}
\caption{\label{fig3} (Color online) Comparison of many-body
calculations with those obtained with the simplified approach in
the case of non-equilibrium initial conditions. For single
electron calculations, we used two different relaxation schemes
shown as insets. This plot was obtained for a 10-site quantum ring
using the same parameter values as in Fig. \ref{fig1}.}
\end{figure}

Next, we have tested the applicability of our approach to highly
excited states. In Fig.~\ref{fig3} we plot the current gnerated in a
10-site ring containing 3 electrons. The difference with previously
discussed calculations is that now we assume that in the initial
moment of time the system is in its highest energy state.  For a
single electron, there are 10 energy states in the 10-site ring. We
made calculations considering different relaxation schemes. Indeed,
there is an arbitrariness in the relaxation state assignment (e.g.,
$V$ operators for the electron which is initially in the 10-th state
-- highest energy state -- can be selected to describe its
relaxation into the first, second or third lowest energy state).
 Fig.~\ref{fig3} displays a very good agreement of the many-body calculation
compared to the results obtained using our single-electron
approach with two different relaxation schemes. Importantly, since
the rates at which electrons relax into their ground states are
the same, the two relaxation schemes lead to the same current,
showing the insensitivity of our general scheme to the details in
the initial state de-population. Also, in the long-time limit the
current is independent of the initial conditions chosen (cf. the
current in Fig.~\ref{fig3} with the current in Fig.~\ref{fig1} at
$t>40$).

\subsection{Precision of the simplified scheme}

\begin{figure}[tb]
\includegraphics[width=8truecm]{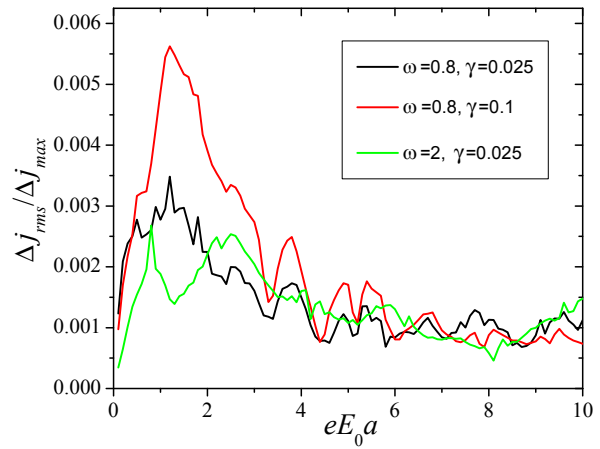}
\caption{\label{fig6} (Color online) RMS of the difference of
currents calculated by many-body and single-electron schemes divided
by the maximum current amplitude within the calculation time as a
function of the electric field amplitude. This plot was obtained for
$N=3$, $M=6$, $\si= 1$, $a=0.1415$nm, $B=10$T and $\tau=100$. The
other calculation parameters are shown in the figure.}
\end{figure}

In order to study the precision of the single-electron scheme, we
calculate the current excited in a 6-site ring containing 3
electrons. Our main observation is that the simplified scheme
provides a very good precision for the whole range of parameters
used in the calculations. We have found that a slightly better
precision is obtained at weak and strong electric fields. This
particular observation is clearly seen in Fig. \ref{fig6} where we
plot the ratio of the RMS of current differences calculated as
\begin{equation}
\Delta j_{rms}=\frac{1}{\tau}\int\limits_0^\tau\sqrt{\left(
j_{mb}-j_{se} \right)^2}dt
\end{equation}
to $\Delta j_{max}=j_{mb}^{max}-j_{mb}^{min}$. Here, $\tau$ is a
sampling period, $j_{mb(se)}$ is the current calculated using
many-body (single-electron) scheme and $j_{mb}^{max(min)}$ is the
maximum (minimum) value of current calculated within the time
interval $[0, \tau]$. A better agreement at weak fields can be
related to the fact that in this situation only the low-energy
states become occupied and the relaxation operators in the
many-electron and single-electron schemes are the same (see Sec.
\ref{analytic} for more arguments). At strong fields, the better
agreement is due to the fact that the electric field term is
dominant in the equations of motion. Fig. \ref{fig6} also
demonstrates that the single-electron scheme precision slightly
depends on simulation parameters and is a better approximation when
dissipation is weaker.

\begin{figure}[tb]
\includegraphics[width=7truecm]{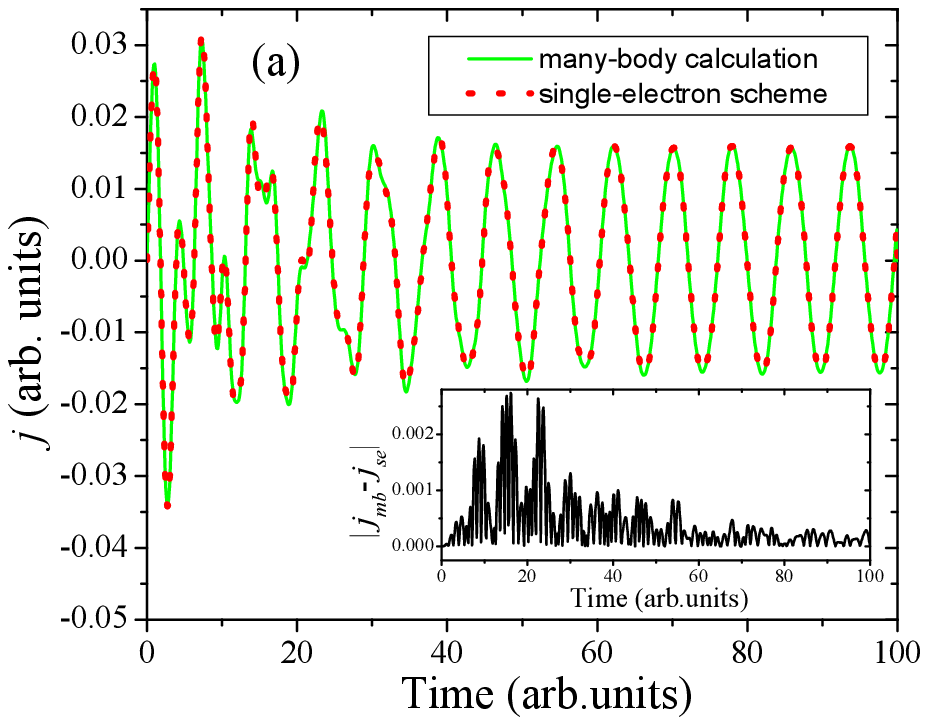}
\includegraphics[width=7truecm]{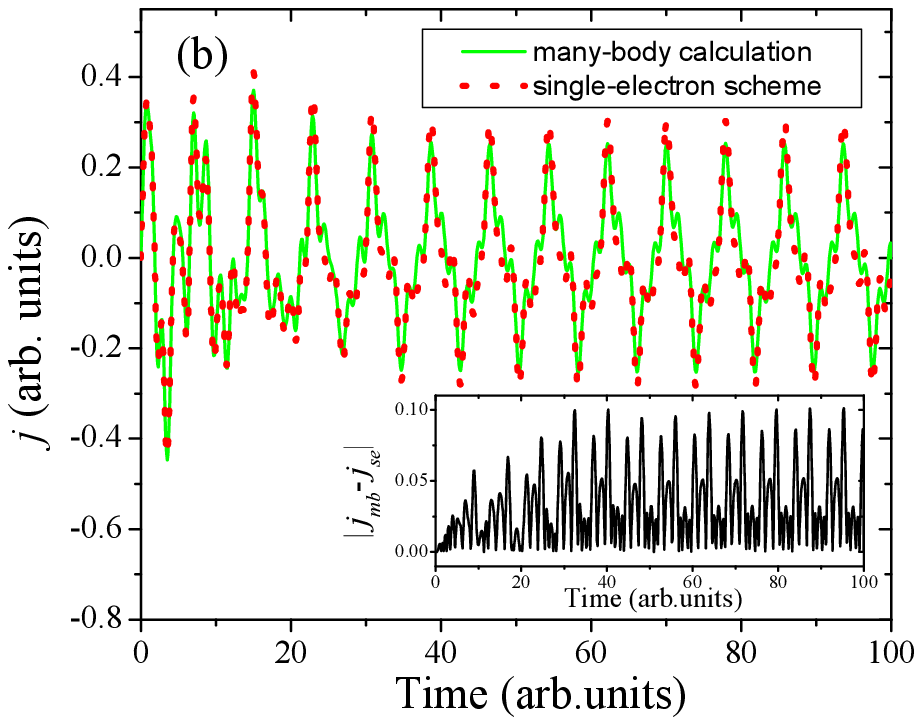}
\includegraphics[width=7truecm]{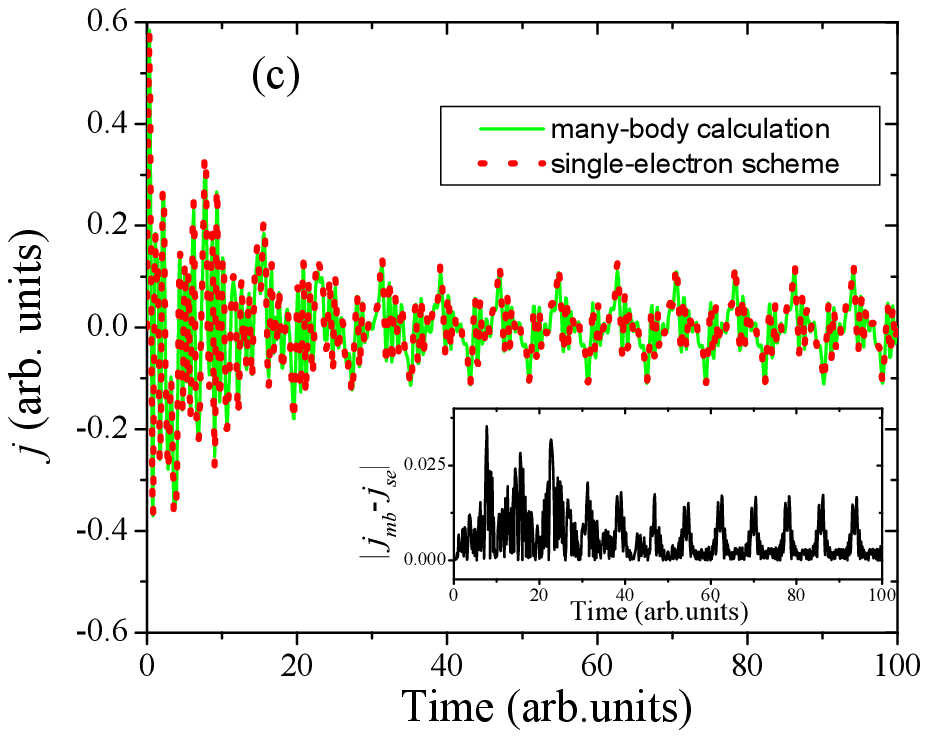}
\caption{\label{fig5} (Color online) Current excited in a 6-site
ring calculated by different approaches as indicated. The
calculation parameters are the same as in Fig. \ref{fig6}. The
electric field amplitude is $eE_0a=0.1$ (a), $1.7$ (b) and $8$ (c).
The insets show the absolute value of many-body ($j_{mb}$) and
single-electron ($j_{se}$) currents difference as a function of
time.}
\end{figure}

Fig. \ref{fig5} presents selected results of our calculations
showing agreement between many-body and single-electron
calculations at several values of the electric field amplitude.
The interesting feature of these results is that at weak driving
fields the single-electron scheme precision is better at longer
times ($t\gtrsim 60$ in Fig. \ref{fig5}(a)), at intermediate
fields the scheme precision is better in the initial time interval
($t\lesssim 20$ in Fig. \ref{fig5}(b)) and at strong fields the
precision is better again at longer times (Fig. \ref{fig5}(c)).

\section{Numerical demonstration: steady state at finite temperatures}\label{numeric1}

\begin{figure}[tb]
\includegraphics[width=8truecm]{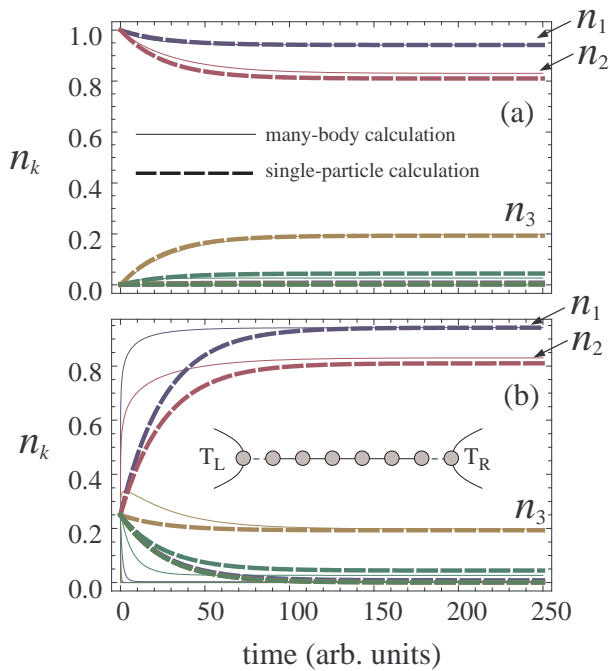}
\caption{\label{compare1} (Color online) Occupation of the different
single-particle energy levels as a function of time for the two
calculation schemes, the full many-body calculation (solid lines)
and the approximate scheme (dashed lines). Initial condition are
either (a) the ground state or (b) a uniformly-occupied state. The
chain length is
 $L=12$, with the parameters $g=1$, $\gamma_0=0.01$, $T_L=0.1$ and
 $T_R=0.4$. }
\end{figure}

In the second numerical example, we study a non-equilibrium system
at finite temperatures. The system of interest is a linear metallic
chain, connected at its two ends to two thermal baths at different
temperatures, $T_L$ and $T_R$, corresponding to the left and right
temperatures (see inset of Fig.~\ref{compare1}(b)).

 The Hamiltonian of the system is given by $\cH=-t \sum_{\langle i,j
\rangle \in L,R,d} \left(\cdag_i c_j + h.c.\right)$  ($t$ is the
hopping integral, which serves as the energy scale, and we have
chosen $t=1$). The master equation now takes the form \beq
\dot{\rho} = -i[\cH,\rho]+\cL_L [\rho]+\cL_R[ \rho]\label{newL}\eeq
where $\cL_{L(R)}$ describes relaxation processes due to the contact
between the left (right) lead with its respective bath at
temperature $T_{L(R)}$. The $V$-operators are given by \beqa
V^{(L,R)}_{kk'}=\sqrt{\gamma^{(L,R)}_{kk'}f^{(L,R)}_D(\e_k)} \bra{k}
\ket{k'} \label{opeq}~~,\eeqa where $f^{(L,R)}_D(\e_k)=1 /\left(
\exp \left (\frac{\e_k-\mu}{k_BT_{L,R}}\right)+1 \right)$ are the
Fermi distributions of the left and right leads, with $\mu$ the
chemical potential. The coefficients \beq \gamma^{(L,R)}_{kk'}=
\left| \psi_k(r)\,\gamma_0\,\psi^*
_{k'}(r)\right|_{r=r_L(r_R)}\label{gamma} \eeq describe the overlap
between the single-particle states $\bra{k}$ and $\bra{k'}$ over the
point of contact $r_{L(R)}$ between the left (right) baths and the
corresponding junction leads. The constant $\gamma_0$ describes the
strength of interactions between the bath and electrons. The
form~(\ref{gamma}) can be derived from first principles by tracing
out the bath degrees of freedom, with the latter formed by a {\em
dense} spectrum of boson excitations (e.g., phonons), which interact
locally with electrons at the edges of the system. Physically, it
corresponds to the experimental situation in which the left (right)
bath induces energy relaxation only between states which reside
predominantly on the left (right) edge of the junction, where the
bath is in contact. The operators~(\ref{opeq}) guarantee that the
system evolves to a global equilibrium if $T_L=T_R$. For $T_L \neq
T_R$ this system is inherently out of equilibrium, and reaches a
steady state which may have, for instance, a non-uniform electron
density~\cite{thermoelectric}, and is thus relevant for experiments
of thermo-power measurements in nano-systems \cite{thermo2}. We
point out that the above model also relaxes the constraint of
Sec.~\ref{numeric} that there is a single relaxation rate for all
relaxation processes.

In Fig.~\ref{compare1} we plot the occupation of the different
single-particle energy levels as a function of time for the two
calculation schemes, the full many-body (solid lines) and the
approximate scheme (dashed lines). The chain length is
 $L=12$, with the parameters $g=1$, $\gamma_0=0.01$, $T_L=0.1$ and
 $T_R=0.4$, and it is occupied by two electrons. We have plotted the dynamics starting from either the
 ground state (Fig.~\ref{compare1}(a)) or a uniform state, where all
 energy levels are equally occupied (Fig.~\ref{compare1}(a)). As
 seen, starting from the ground state (Fig.~\ref{compare1}(a)) there is excellent agreement
 between the two schemes both in the transient dynamics and in the
 steady state. On the other hand, if we start from an excited state
 (Fig.~\ref{compare1}(b)) then the transient dynamics exhibit slight
 differences between the exact and approximate scheme. The steady
 state is, naturally, the same with either initial conditions. Similar calculations with
 different parameters have yielded similar results.

\begin{figure}[tb]
\includegraphics[width=6truecm]{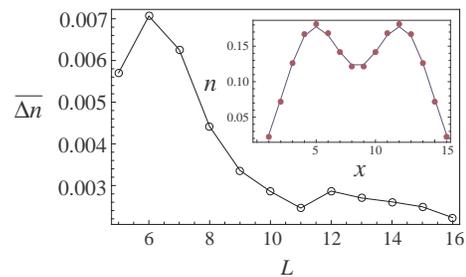}
\caption{\label{compare2} (Color online) Difference in the local
density $\overline{\D n}$  (averaged over the entire chain) as a
function of system length at steady state. The numerical
parameters are the same as in Fig.~\ref{compare1}. Inset : local
density along a $L=16$ chain, calculated using the exact scheme
(points) and approximate scheme (solid line).}
\end{figure}

In order to study the accuracy of the approximation also in the
present example, we calculate the difference in the local density
between the two schemes, $\D n_i=| n_{i,mb}-n_{i,sp}|$, at steady
state. Here, $n_{i,mb(sp)}$ is the local density ($n_i=\sum_k
|\psi_k(i)|^2 \rho_{kk}$) at the $i$-th site, calculated with the
many-body (single-particle) scheme. In Fig.~\ref{compare2} we plot
$\D n$ (averaged over the entire chain), for the same parameters
as in Fig.~\ref{compare1} for different chain lengths
$L=5,6,...,16$. We find that as the system becomes larger the
approximation improves (the relative deviation for the larger
systems is less than $3 \%$). The reason for the improvement of
the approximation with increasing length stems from the fact that
as the system becomes larger, the single-particle occupations of
the many-body system become closer and closer to a true broadened
Fermi distribution. In the inset of Fig.~\ref{compare2} we plot
the local density along a $L=16$ chain, calculated using the exact
scheme (points) and approximate scheme (solid line), showing the
excellent agreement between the two.

\section{Analytic justification} \label{analytic}
We now provide an analytical argument for the validity of our {\em
ansatz}~, which is summarized in
Eqs.~(\ref{theorem}-\ref{V_effective}) for $T=0$. In order to do
so we start from the definition of an auxiliary single-particle
density matrix (SPDM) from the many-body one. We then evaluate its
equation of motion by summing up the many-body degrees of freedom,
and study the structure of the equations. We in fact find that
this SPDM can be approximately written as sum of single-particle
density matrices obeying equations of motion with specific bath
operators, thus validating our {\em ansatz}. We do this for finite
temperatures, and show that the result leads to the $T=0$ form for
the relaxation operators used in the numerical calculations.

Let us define the following SPDM \beq
\rho(t)=\sum_{kk'}\rho_{kk'}(t) \bra{k} \ket{k'}\label{singlerho}~.
\eeq The matrix elements are derived from the many-body DM by \beq
\rho_{kk'}=\Tr \left( \cdag_k c_{k'} \rho_M\right). \eeq We show
below that $\rho(t)$ can be approximated as
\begin{equation}
\rho(t)\simeq \sum_{j} \rho^{(j)}(t),\label{singlerho1}
\end{equation} where $\rho^{(j)}$ are the single-particle density
matrices entering Eq.~(\ref{theorem}).

The time evolution of the SPDM is determined by \beq
\dot{\rho}_{kk'}=\frac{\mathrm{d}}{\mathrm{d}t}\Tr \left( \cdag_k
c_{k'} \rho_M\right) = \Tr \left( \cdag_k c_{k'}
(-i[\cH,\rho_M]+\cL \rho_M)\right). \label{trace}\eeq One can now
perform the trace exactly using Wick's theorem. The relaxation
operators $V_{nn'}$ defined below Eq. (\ref{Lindbladian})
generally involve up to $M$ creation and $M$ annihilation
operators. Therefore, it is not practical to use them in
analytical calculations. We instead consider $V$ operators of a
commonly used \cite{Gebauer} simplified form $V_{k k'}=\left(
\gamma_{kk'}\right)^{\half} \cdag_k c_{k'}~~, k \neq k'$. It is
clear that when excitation of the system is weak, and highly
excited states are almost unpopulated, the physical effect caused
by both operators is nearly the same. Note, however, that taking
this form for the $V$-operators (which excludes direct relaxation
of highly-excited states into the ground state) does not lead to a
reduction in the number of equations needed to be solved, since
the equations remain fully coupled (put it differently, the
Lindbladian operator cannot be subdivided into blocks).

\subsection{Diagonal elements}

We start by deriving equations of motion for the diagonal elements
of the SPDM. For the sake of simplicity, let us assume that the
system Hamiltonian is time-independent and diagonalized. Then, it
is easy to find that the equations of motion for the diagonal
elements of the SPDM are \cite{Huang} \beqa \dot{\rho}_{kk} &=&
-\half\sum_{k'\not{=}k} \gamma_{kk'}\rho_{kk}+\half
\sum_{k'\not{=}k} \gamma_{k'k}\rho_{k'k'} + \nonum & &~~+
 \half \rho_{kk}\sum_{k'\not{=}k}(\gamma_{k'k}-\gamma_{kk'}) \rho_{k'k'}.\label{rhodot}\eeqa

Let us examine Eq.~(\ref{rhodot}) by making two assumptions:~(i) the
coefficients are only a function of the first index, i.e.,
$\gamma_{kk'}=\gamma_{k'}$, and (ii) the third (non-linear) part on
the RHS of Eq.~(\ref{rhodot}) is negligible and is set to zero.
Within these assumptions, and noting that by definition
$\sum^M_{k=1}\rho_{kk}=N$, one obtains the equation \beq
\dot{\rho}_{kk}=-Z\rho_{kk}+\gamma_k(N-\rho_{kk}),
\label{simple}\eeq where $Z=\sum_{k'\neq k}\gamma_{k'}$. Solving
this equation yields \beq \rho_{kk}(t)=\left(
\rho_{kk}(0)-\frac{\gamma_k N}{\gamma_k+Z}\right)
\e^{-(Z+\gamma_k)t}+\frac{\gamma_k N}{\gamma_k+Z}.\eeq For a Fermi
system, the long-time limit of the SPDM should be $\rho_{kk}(t \to
\infty)=f_D(\e_k)$, where $f_D(\e_k)=1/(1+\exp((\e_k-\mu)/k_BT)$ is
the Fermi-Dirac distribution. It follows directly that in order to
satisfy this long-time limit, the coefficients must be chosen such
that $\gamma_k=\gamma f_D(\e_k)$.

We now turn back to the third, non-linear part in the RHS of
Eq.~(\ref{rhodot}). Keeping in mind the definition for $\gamma_{k}$,
this part now reads $\gamma \rho_{kk}\sum_{k'\neq
k}(f_D(\e_k)-f_D(\e_{k'}))\rho_{k'k'}$. In the long-time limit, as
$\rho_{kk}$ approach their equilibrium values, and at zero
temperature, one can consider two possibilities. In the first, both
$k$ and $k'$ lie below or above the Fermi surface. In this case,
$f_D(\e_k)-f_D(\e_{k'})\approx 0$ and the non-linear part vanishes.
If, on the other hand, either $k$ or $k'$ lie below the Fermi
surface and the other above it, then indeed $f_D(\e_k)-f_D(\e_{k'})
\neq 0$. However, in that case either $\rho_{kk}\approx 0 $ or
$\rho_{k'k'}\approx 0 $. Thus, in the low temperature long-time
limit, the third term on the RHS of Eq.~(\ref{rhodot}) is
negligible, which means that our assumption (ii) above is justified.

Extending this conclusion to finite temperatures and to all times,
we end up with a simple equation for the diagonal elements of the
SPDM,
 \beq \rho_{kk}=-\gamma \sum_{k'\neq k} f_D(\e_{k'}) \rho_{kk}
+\gamma \sum_{k'\neq k} f_D(\e_k) \rho_{k'k'}.\eeq Simple algebra
reveals that these equations are equal to those obtained from
applying the Lindbladian operator, Eq.~(\ref{Lindbladian}), to the
SPDM, with the $V$-operators having the form \beq
V_{kk'}=\sqrt{\gamma f_D(\e_k)} \bra{k} \ket{k'}
\label{V-operators} ~~,\eeq which is a particular case of the
operators~(\ref{opeq}), thus justifying their structure. We thus
propose that the SPDM evolves according to Eq.~(\ref{lind_eq1})
and~(\ref{Lindbladian}), with the Lindblad operator given in terms
of Eq.~(\ref{V-operators}).

The equations of diagonal SPDM elements can be derived differently.
Since $\rho(t)\simeq \sum_{j} \rho^{(j)}(t)$, using Eq.
(\ref{lind_eq2}) with the single-electron $V$-operators in the form
\beq V^j_{kk'}=\del_{kj}(1-\del_{kk'})\sqrt{\gamma
f_D(\e_k)}\bra{j}\ket{k'} , \label{V_2}\eeq we can obtain a set of
equations which is the same as Eq. (\ref{rhodot}). This
demonstration clearly shows a similarity of our single-electron and
many-body approaches. Note, that the definition (\ref{V_2})
coincides with Eq.~(\ref{V_effective}) at $T=0$. Moreover, while
there is no {\em a priori} justification for neglecting the
non-linear terms, the numerical calculations of the previous
sections show that it is an excellent approximation for
non-interacting systems.

Let us also point out that the equations for the diagonal and
off-diagonal parts of the SPDM are completely decoupled (this result
is exact). Therefore, if one is interested in the time-dependent
expectation value of an operator that commutes with the Hamiltonian,
or only in the steady-state (where the off-diagonal elements vanish)
our {\em ansatz} reduces the computational effort to a single $M
\times M$ equation for the diagonal elements of the SPDM.

\subsection{Off-diagonal elements}
The off-diagonal elements of the density matrix are needed to
calculate, e.g., local currents or densities in a non-equilibrium
situation of an excited system (as in the numerical examples of
Sec.~\ref{numeric}). As stated above, if only the diagonal elements
are of interest, SPDM calculations with the $V$-operators in their
especially simple form (Eq.~\ref{V-operators}) can be used. If the
off-diagonal elements are important, calculations using
single-electron matrices $\rho^j$ with relaxation operators given by
Eq.~\ref{V_2} have to be performed.

\begin{figure}[t]
\includegraphics[width=9truecm]{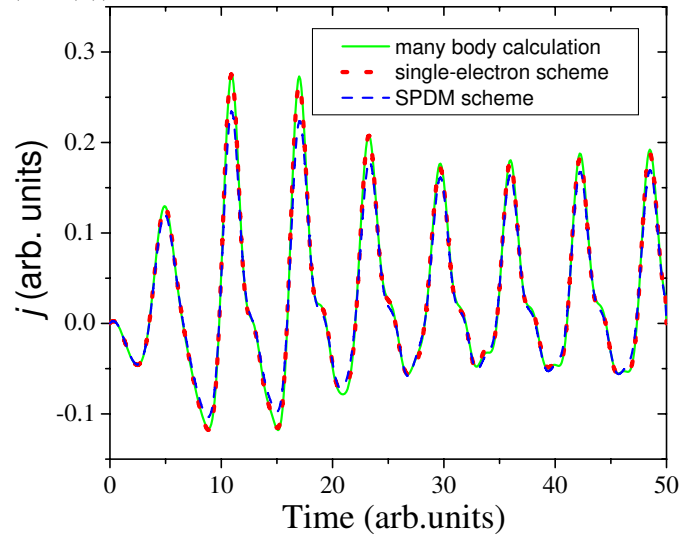}
\caption{\label{fig4} (Color online) Current excited in a 10-site
ring calculated by three different methods as indicated in the
figure. This plot was obtained using the same parameter values as in
Fig. \ref{fig1} except $\omega=1$.}
\end{figure}

In order to understand why single-electron calculations are needed
(or why SPDM does not provide the best results in all cases), we
study the equation of motion for the off-diagonal elements of the
exact Lindblad operator. Using Eq.~(\ref{trace}) and $V_{kk'}$
operators defined below Eq.~(\ref{trace}) one finds \beqa (\cL
\rho)_{kk'} &=& -\half \sum_{k'' \neq k,k'}(\gamma_{k''
k'}+\gamma_{k'' k})(1-\rho_{k'' k''}) \rho_{kk'}- \nonum  &&
~~-\half\sum_{k'' \neq k,k'}(\gamma_{kk''}+\gamma_{k' k''})\rho_{k
k'} \rho_{k''k''}. \label{offdiag1} \eeqa Again we make the
substitution $\gamma_{kk'}=\gamma f_D(\e_{k'})$, and consider for
simplicity the system at zero temperature. By assuming
$\rho_{kk}\approx f_D(\e_k)$ we find that the first element of the
LHS in Eq.~(\ref{offdiag1}) is negligible, and one is left with \beq
(\cL \rho)_{kk'} \approx -\frac{\gamma}{2} \sum_{k'' \neq k,k'}
\rho_{k''k''} \rho_{kk'} =-\frac{\gamma}{2}
(N-\rho_{kk}-\rho_{k'k'}) \rho_{kk'}. \label{offdiags} \eeq If one
uses SPDM calculations to study the off-diagonal elements, then one
finds that $ (\cL \rho)_{kk'}$ does not depend on
$\rho_{kk},~\rho_{k'k'}$ at all. However, within the single-electron
scheme this separation can not be made, and the dynamics of the
off-diagonal elements are better captured. This can be seen in the
numerical example by comparing the exact many-body calculation with
the approximate calculation using both Eq.~(\ref{singlerho1}) and
the SPDM Eq.~(\ref{singlerho}). This is shown in Fig.~\ref{fig4},
where a comparison between the three methods is shown. As seen in
the figure, the agreement between all schemes is good in general,
with substantial differences arising only at the maxima and minima
of the current. At these points, the single-particle scheme
[Eq.~(\ref{singlerho1})] is closer to the many-body calculation than
the SPDM method [Eq.~(\ref{singlerho})].

\section{Summary} \label{summary} We have proposed an order-$N$ scheme to investigate
the dynamics of $N$ non-interacting electrons coupled to one or more
baths, and justified it analytically by examining and tracing the
full many-body calculation. The main idea is to reduce the equation
of motion for the many-body system to a set of effective
single-electron equations (Eq.~(\ref{lind_eq2})) where both Fermi
statistics and dissipation are taken into account via a specific
form of relaxation operators (Eq.~(\ref{V_effective}) at $T=0$;
Eq.~(\ref{V_2}) for $T \neq 0$). We have numerically demonstrated
that the proposed method is in excellent agreement with the exact
many-body calculation by studying two example. The first example is
a system of tight-binding rings at zero temperature, driven out of
equilibrium by external radiation. The second example is a linear
chain connected at its end to two heat baths held at different
temperatures.

Since, even for non-interacting electrons the inclusion of the Pauli
exclusion principle is nontrivial for open quantum
systems~\cite{transp}, we believe our scheme can be used in systems
where interactions play a relatively minor role such as in graphene
\cite{Graphene}, quantum point contacts \cite{QPC}, etc.
Nevertheless, while the above examples did not include
electron-electron interactions, the latter may be included within
the framework of stochastic time-dependent current-density
functional theory~\cite{DiVentra2007}, where the interacting
many-body problem in the presence of environments is mapped into an
effective single-particle problem in the presence of the same
environments. Our {\em ansatz} thus provides a good starting point
to solve the corresponding equations of motion with a computational
cost that scales only linearly with the number of particles. Such a
project is currently underway.

\acknowledgements
We thank S. Saikin for fruitful discussions. This work was funded by
the Department of Energy grant DE-FG02-05ER46204.


\begin{thebibliography}{}
\bibitem{Feynman}
R. P. Feynman and F. L. Vernon, Ann. Phys. (N.Y.) {\bf 24}, 118
(1963).

\bibitem{Legget}
A. O. Caldeira and A. J. Legget, Ann. Phys. (N.Y.) {\bf 149}, 374
(1983).

\bibitem{Weiss}
For a comprehensive review, see, e.g., U. Weiss, {\it Quantum
Dissipative Systems}, Series in Modern Condensed Matter Physics,
Vol. 10 (World Scientific, Singapore, 2006).

\bibitem{Max}
M. Di Ventra, {\it Electrical Transport in Nanoscale Systems}
(Cambridge University Press, 2008).

\bibitem{Bush1} N. Bushong \etal, Nano Letters
{\bf 5}, 2569 (2005).

\bibitem{Redfield}
 A. G. Redfield, IBM J. Res. Dev. {\bf 1}, 19 (1959).

\bibitem{vancamp} N. G. Van Kampen, {\it Stochastic Processes in Physics and Chemistry}
(North Holland, Amsterdam, 2001), 2nd ed.

\bibitem{lind} G. Lindblad, Commun. Math. Phys. {\bf 48}, 119 (1976).

\bibitem{Louisell}
W. H. Louisell, {\sl Quantum Statistical Properties of Radiation}
(Wiley-Interscience, 1990).

\bibitem{Breuer}
An excellent review on the on the use of both the density-matrix methods
and the stochastic Schroedinger equation in various problems may be found in, e.g.
H. -P. Breuer and F. Petruccione, {\it The Theory of Open Quantum Systems}  (Oxford University Press,
Oxford 2002).

\bibitem{DiVentra2007}
M. Di Ventra and R. D'Agosta, Phys. Rev. Lett. {\bf 98}, 226403
(2007); R. D'Agosta and M. Di Ventra, cond-mat/08053734.

\bibitem{Neil}
N. Bushong and M. Di Ventra, cond-mat/0711.0762 (2007).

\bibitem{stochN}
For the stochastic Schr\"odinger equations we
need to solve for $C_N^M-1$ elements of the state vector but average
over an amount, call it $m$, of different realizations of the
stochastic process.

\bibitem{perpier} Yu. V. Pershin and C. Piermarocchi, Phys. Rev. B {\bf 72}, 245331
(2005); Phys. Rev. B {\bf 72}, 125348 (2005).

\bibitem{Nobusada} K. Nobusada and K. Yabana, Phys. Rev. A {\bf 75}, 032518 (2007).

\bibitem{prec} Note that for our choice of parameters the continuity
equation is satisfied to a high degree of accuracy.

\bibitem{thermoelectric}
Y. Dubi and M. Di Ventra, cond-mat/08051415.

\bibitem{thermo2}
See, e.g. B. Ludoph and J. M. van Ruitenbeek, \prb {\bf 59}, 12290
(1999), and other references in Ref.~\onlinecite{thermoelectric}.


\bibitem{Gebauer}
See, e.g., R. Gebauer and R. Car, \prb {\bf 70}, 125324 (2004).

\bibitem{Huang}
C. F. Huang, and K.-N. Huang, Chinese J. Phys. {\bf 42}, 221 (2004).

\bibitem{transp}
L. Bonig and K. Schonhammer, \prb {\bf 47}, 9203 (1993).

\bibitem{Graphene}
H. P. Dahal \etal,  \prb {\bf 74}, 233405 (2006); H. P. Dahal \etal,
cond-mat/0712.2836 (2007).

\bibitem{QPC}
F. A. Maao and L. Y. Gorelik, Phys. Rev. B {\bf 53} 15885 (1996).



\end{thebibliography}
\end{document}